\documentclass[aps,prd,twocolumn,showpacs,showkeys,superscriptaddress]{revtex4-2}
\usepackage{latexsym}
\usepackage{amssymb}
\usepackage{amsmath}
\usepackage{amscd}
\usepackage{amsthm}
\usepackage{graphicx}
\usepackage{textcomp}
\usepackage{colortbl}
\usepackage[colorlinks]{hyperref}
\usepackage[font={footnotesize,it}]{caption}
\usepackage{multirow}
\usepackage[T1]{fontenc}
\usepackage{ae,aecompl}
\usepackage{subcaption}
\begin{document}

\title{Interacting Dark Sectors in Anisotropic Universe: Observational Constraints and $H_{0}$ Tension}
\author{Hassan Amirhashchi}
\email[]{h.amirhashchi@mhriau.ac.ir}
\affiliation{Department of Physics, Mahshahr Branch, Islamic Azad University, Mahshahr - 6351977439, Iran}
\author{Anil Kumar Yadav}
\email[]{abanilyadav@yahoo.co.in}
\affiliation{Department of Physics, United College of Engineering and Research,Greater Noida - 201310, India}
\author{Nafis Ahmad}
\email[]{nafis.jmi@gmail.com}
\affiliation{Department of Physics, College of Science, King Khalid University, P.O. Box 9004, Abha 61413, Saudi Arabia}
\author{Vikrant Yadav}
\email[]{vikuyd@gmail.com}
\affiliation{School of Basic and Applied Sciences, Raffels University, Neemrana - 301705, Rajasthan, India}
\begin{abstract}
We present observational constraints on the coupling between dark components of anisotropic Bianchi type I universe. We assume interaction between dark matter and dark energy, and split the continuity equation with the inclusion of interaction term $\Gamma$ in two scenarios: (i) when coupling between dark components is constant and (ii) when it is a function of redshift ($z$). We utilize Metropolis-Hastings algorithm to perform Monte Carlo Markov Chain (MCMC) analyses of the models by using observational Hubble data from cosmic chronometers (CC) technique, cosmic microwave background (CMB), baryon acoustic oscillation (BAO), Pantheon compilation of Supernovae type Ia (SNIa), and a Gaussian prior on the Hubble parameter $H_{0}$. We find that the combination of all databases plus $H_{0}$ prior gives stringent constraints on the coupling parameter, viz., $-0.001<\delta<0.041$ in the constant coupling model and $-0.042<\delta_0<0.053$ in varying coupling model, both at 68\% CL. In general, for both models, we find $\omega^{X}\approx -1$ and $\delta(\delta_{0})\approx 0$, which indicate that observational data favor $\Lambda$CDM model with null interaction among the dark sector components. In the constant coupling model, our estimations show that  $(H_{0}=73.9^{+1.5}_{-0.95}, \delta=0.023^{+0.017}_{-0.024})$. This result is interesting because the previous works show that such a high value of Hubble constant requires the significant value of coupling parameter $\delta$. 
\end{abstract}


\maketitle
\section{Introduction}
\label{sec:intro}
Type Ia supernovae (SNIa) and cosmic microwave background (CMB) observations have now confirmed that we live in an accelerating expanding universe \cite{Perlmutter/1998,Riess/2004,Caldwell/1998,Huang/2006,Tegmark/2004}. However, it is known that the universe expansion has already changed from decelerating to accelerating phase at a certain redshift called \textgravedbl transition redshift\textacutedbl. This phase transition is due to an unknown mechanism changing the sign of the universal deceleration parameter $q(z)$. In other words, the actual cause of this late time acceleration which acts against the gravitational force, is still unknown. To describe the kinematics and fate of current universe, one has to assume either presence of an energy source in the context of General Theory of Relativity (GR) or a modification of GR by introducing some additional terms in Ricci scalar. In the context of GR, the present acceleration of the universe is commonly
attributed to some exotic form of energy designated dark energy. The recent result obtained from Plank collaboration indicates that about 70\% of the total energy of the universe is in the form of dark energy \cite{Ade/2014,Ade/2016}. A possible candidate for dark energy is that it is vacuum energy of some kind, equivalent to a cosmological constant $\Lambda$.  But the inferred vacuum energy density is many orders of magnitude smaller than simplistic estimates from quantum field theory, and the approximate equality of matter and vacuum energy densities (in order) today remains an unexplained coincidence \cite{Wienberg/1989,Sahni/2000}. Therefore, some dynamical cosmological models like quintessence \cite{Martin/2008}, phantom \cite{Caldwell/2003}, chaplygin gas \cite{Bento/2002} and interacting \cite{Zimdahl/2004,Zhang/2005a,Zhang/2005b,Setare/2011,Amirhashchi/2014} dark energy scenarios have been proposed from time to time. Keeping in mind that the dark components of universe play a major role in driving the late time acceleration of the universe, the observational evidence for dark components motivates the study of the coupling between dark components of Bianchi I (Henceforth BI) universe.

The standard cosmological model assumes large scale isotropy and homogeneity, but in the literature, several investigations beyond the standard model exist. According to analysis of WMAP data \cite{Hinshaw/2009,Hinshaw/2003,Jaffe/2005}, a small amount of anisotropy may be possible in the universe. Therefore, a more general and realistic study of evolution of universe may require modification in terms of inclusion of anisotropy in the structure of universe that leads to the scope of Bianchi morphology \cite{Jaffe/2006,Campanelli/2006,Campanelli/2007}. Recent observations such as CMB experiment \cite{Eriksen/2007} and LSS observations indicate that there is a tiny fluctuation in the intensity of CMB coming from different directions in the sky. However, to handle the issue of anisotropy, Bianchi type models become natural choice of the cosmologists \cite{Koivisto/2008,Battye/2009,Campanelii/2009,Akarsu/2010,Yadav/2011,Kumar/2011,Hassan/2011}. Ellis \cite{Ellis/2006} has already pointed out that although the observed universe seems to be almost isotropic on large scales, the early and/or very late universe could be anisotropic. Moreover, Goliath and Ellis \cite{Goliath/1999} have shown that some Bianchi models isotropise due to inflation. Saadeh et al \cite{Saadeh/2016} have recently conducted a general test of isotropy using cosmic microwave background temperature and polarization data from Planck. It is worth noting that the BI cosmological model is the general form of Friedman-Lema\^itre-Robertson-Walker (FRLW) model, where the spatial isotropy is relaxed. Among the several Bianchi type models, BI cosmological model has attracted more attention due to its fundamental properties: it has more degrees of freedom with respect to FLRW characterized by Lie groups and it recovers isotropic scenario as special case and permits a small amount of anisotropy. This small amount of anisotropy may affect the physical behavior of the universe in early times of evolution. The spatial section of BI space-time is flat but the expansion rates are direction dependent. In the recent past numerous anisotropic cosmological models have been constructed to study the different aspects of the accelerating universe 
\cite{Singh/2011,Yadav/2012,Amirhashchi/2017,Muharlyamov/2019,Hassan/2019,Hassan/2018,Mishra/2017,Mishra/2019,Yadav/2021PDU,Yadav/2021PRD}. In 2015, Bolotin et al \cite{Bolotin/2015} have studied coupling between dark matter and dark energy. This study reveals that the kinematics and fate of two fluid interacting universe can differ significantly from the standard cosmological model.

In the present universe, it is reasonable to consider the gravitational interaction between dark matter and dark energy due to major contributions of relative densities of these components. In fact, this scenario leads to a solution to the coincidence problem \cite{Cimento/2003, Setare/2007} and also provides a natural way to detect dark energy. It is worth mentioning that some observations \cite{Bertolami/2007, Le Delliou/2007, Berger/2006, Salvatelli/2014, Di Valentino/2017, Yang/2018, Pan/2019} have already supported the possibility of such an interaction. Moreover, recent studies show that such an interacting scenario results in alleviating the two known tensions of modern cosmology, i.e., $H_{0}$ \cite{Valentino/2015, Valentino/2016, Renk/2017, Arenas/2018, Yang/2018, Poulin/2019, Khosravi/2019, Yang/2019a, Yang/2019b, Vattis/2019} and $\sigma_{8}$ \cite{Kumar/2019, Pourtsidou/2016, An/2018}. Wetterich \cite{Wetterich/1995} has investigated a scalar field cosmological model by taking into account the coupling between gravity and a scalar field with exponential potential. In Refs. \cite{Amendola/2003,Hoffman/2003,Chimento/2003,Amendola/2004,Campo/2005,Guo/2005,Jesus/2008}, the authors have investigated the cosmological models by assuming dynamical form of dark energy and its interaction with dark matter. Note that the scalar field dark energy models are minimally coupled with gravity and do not allow non-minimal interaction of field to the background matter. Since the exact nature of either dark energy or dark matter is still unknown, one can not exclude the coupling between these dark components of universe. Some important applications of interaction in dark sector of universe are given in Refs. \cite{Huey/2006,Das/2006,Chimento/2009,Binder/2006,Binder/2006,Farrar/2004,Guo/2007}. Recently Kumar et al. \cite{Kumar/2019} have studied the interaction scenario of dark sector of universe with Planck-CMB, KiDS and HST data. This analysis shows that there is strong statistical support from joint Planck-CMB, KiDS and HST data for an interaction between dark energy and dark matter. In Martinelli et al \cite{Martinelli/2019}, the authors have tested an interacting scenario between vacuum energy and geodesic cold dark matter by using combined CMB data from Planck \cite{Ade/2016,Ade/2016a}, BAO, redshift space distortion and SN Ia data to constrain various parametrizations of coupling parameter. In this paper, we confine ourselves to constraining the coupling between dark matter and dark energy of BI universe.  

The paper is organized as follows: Section \ref{sec:2} deals with the model and basic mathematical formalism. In Section \ref{sec:3}, we derive a general differential equation for interacting DM-DE in the scope of BI space-time and solve it analytically. In subsections \ref{subsec:1} \& \ref{subsec:2}, we derive analytical solution for constant and varying coupling models respectively. We introduce the computational method which has been used in this paper to fit model parameters to data by a numerical MCMC analysis in \ref{data}. Section \ref{results} deals with the results of MCMC analyses of the models with the data. Finally, in section \ref{summary}, we summarize our findings.
\section{Theoretical model and Basic equations}
\label{sec:2}
The Bianchi type I space time reads 
\begin{equation}
\label{eq1}
ds^{2} = -dt^{2} + A^{2}dx^{2}+B^{2}dy^{2}+C^{2}dz^{2},
\end{equation}
where $\{A(t),~ B(t)~ \& ~C(t)\}$ are scale factors along $x$, $y$ and $z$ axis respectively. Thus the average scale factor is given by $a = (ABC)^{\frac{1}{3}}$.\\
The Einstein's field equation is 
\begin{equation}
\label{eq2}
R_{ij} -\frac{1}{2}R g_{ij}  = -8\pi G \left(T_{ij}^{m} + T_{ij}^{X}\right).
\end{equation}
Here, $T_{ij}^{m}$ and $T_{ij}^{X}$ are the energy momentum tensors of the dark matter and dark energy respectively (note that in Section III the matter and dark energy densities will be taken to evolve as power laws with cosmic scale factor $a$, but in this section
their evolution is permitted to be arbitrary), given by\\

$T_{ij}^{m} = diag[-\rho^{m},p^{m},p^{m},p^{m}]$\;\;\\

\&\\

$T_{ij}^{X} = diag[-\rho^{X},p^{X},p^{X},p^{X}]$.\;\;\\

For metric (\ref{eq1}), the field equation (\ref{eq2}) yields the following set of differential equations:
\begin{equation}
\label{eq3}
 \frac{\ddot{B}}{B}+\frac{\ddot{C}}{C}+\frac{\dot{B}\dot{C}}{BC} = -8\pi G\left(p^{m}+p^{X}\right),
\end{equation}
\begin{equation}
\label{eq4}
 \frac{\ddot{C}}{C}+\frac{\ddot{A}}{A}+\frac{\dot{A}\dot{C}}{AC} = -8\pi G\left(p^{m}+p^{X}\right),
\end{equation}
\begin{equation}
\label{eq5}
 \frac{\ddot{A}}{A}+\frac{\ddot{B}}{B}+\frac{\dot{A}\dot{B}}{AB} = -8\pi G\left(p^{m}+p^{X}\right),
\end{equation}
\begin{equation}
\label{eq6}
 \frac{\dot{A}\dot{B}}{AB}+\frac{\dot{B}\dot{C}}{BC}+\frac{\dot{C}\dot{A}}{CA} = 8\pi G \left(\rho^{m}+ \rho^{X}\right).
\end{equation}
The Bianchi identity $G_{ij}^{;j} = 0$ leads $T_{ij}^{;j} = 0$, therefore the equation of continuity for metric (\ref{eq1}) in connection with equation (\ref{eq2}) reads as
\begin{equation}
\label{eq7}
{\dot{\rho}^{m}}+3H\rho^{m}+{\dot{\rho}^{X}}+3H\left(\rho^{X}+ p^{X}\right) = 0.
\end{equation}
 
Solving equations (\ref{eq3}) - (\ref{eq5}), we obtain the following relation among the directional scale factors:\\
\begin{equation}
\label{eq8}
\frac{B}{A} = d_{1}\exp\left(\int \frac{x_{1}}{ABC}dt\right),
\end{equation}

\begin{equation}
\label{eq9}
\frac{C}{A} = d_{2}\exp\left(\int \frac{x_{2}}{ABC}dt\right),
\end{equation}

\begin{equation}
\label{eq10}
\frac{C}{B} = d_{3}\exp\left(\int \frac{x_{3}}{ABC}dt\right),
\end{equation}
where $d_{1},~ d_{2},~ d_{3},~x_{1},~x_{2},~x_{3}$ are arbitrary constants of integration. Without loss of generality, we assume  that $d_{1} = d_{2} = d_{3} = 1$ and $x_{1} = -x_{2} = k$, $x_{3} = -x_{1}+x_{2}$. Thus after some algebra, equations (\ref{eq8})-(\ref{eq10}) lead the following relations among directional scale factor:\\ 
 
$B = AD$, $C = AD^{-1}$ and $D = \exp\left[\int\frac{k}{ABC}dt\right]$, where $D = D(t)$ is defined as the anisotropic term. In this way, average scale factor $a$ is obtained as\\
\begin{equation}
\label{eq11}
a = (ABC)^{\frac{1}{3}} = A.
\end{equation}
Now, the Hubble's parameter is given by 
\begin{equation}
\label{eq12}
H = \frac{\dot{a}}{a} = \frac{1}{3}\left(\frac{\dot{A}}{A}+\frac{\dot{B}}{B}+\frac{\dot{C}}{C}\right) = \frac{\dot{A}}{A}.
\end{equation}

Finally Friedmann equation (\ref{eq6}) for the anisotropic BI universe can be recast as follows:
\begin{equation}
\label{eq13}
H^{2} = \frac{\dot{a}^{2}}{a^{2}} = \frac{8\pi G}{3}\left(\rho^{m}+\rho^{X}+ \frac{k^{2}}{8\pi G a^{6}}\right),
\end{equation}
where the third term on RHS relates the anisotropy of space time.\\

We assume the interacting scenario between dark matter and dark energy. Therefore, we can split equation (\ref{eq7}) as follows: 
\begin{equation}
\label{eq14}
{\dot{\rho}^{m}}+3H\rho^{m} = \Gamma \rho^{X},
\end{equation} 
\begin{equation}
\label{eq15}
{\dot{\rho}^{X}}+3H\left(\rho^{X}+ p^{X}\right) = -\Gamma \rho^{X}, 
\end{equation} 
where $\Gamma$ is the interaction term and in connection with scalar field dark energy models it is defined as $\Gamma = Q\dot{\phi}$ with $Q$ is a constant that characterizes the strength of the coupling and $\phi$ denotes a scalar field \cite{Amendola/2000,Gumjudpai/2005}. In this paper, our approach is different from the scalar field dark energy models such that origin of dark energy does not associate with the scalar field \cite{Guo/2007}. For this purpose, we define $\delta = \frac{\Gamma}{H}$ with $\delta > 0$, and it implies a transfer of energy from dark matter to dark energy and vice-versa. \\

\section{General Solution}
\label{sec:3}
In this section, we obtain a general solution for interacting DM-DE in the scope of BI space-time. The first integral of Equation (\ref{eq15}) leads to 
\begin{equation}
\label{eq16}
\rho^{X}=\rho_{0}a^{-3(1+\omega^{X})}\exp \left[-\int \delta d(ln a)\right],
\end{equation}
where $\omega^{X} = \frac{p^{X}}{\rho^{X}} = \mbox{constant}$ is the dark energy equation of state parameter (EOS). Following Dalal et al \cite{Dalal/2001}, we assume that the dark energy and dark matter are coupled with following relation:

\begin{equation}
\label{eq17}
\frac{\rho^{m}}{\rho^{X}} = A^{-1}a^{-\eta}, ~~~A\equiv\frac{\rho^{X}_{0}}{\rho^{m}_{0}}=\frac{\Omega^{X}_{0}}{\Omega^{m}_{0}}.
\end{equation}
This equation gives
\begin{equation}
\label{eq18}
\rho^{X}=\frac{Aa^{\eta}}{1+Aa^{\eta}}(\rho^{tot}-\rho^{\sigma});~~\rho^{m}=\frac{1}{1+Aa^{\eta}}(\rho^{tot}-\rho^{\sigma}),
\end{equation}
where $\rho^{tot}=\rho^{X}+\rho^{m}+\rho^{\sigma}$, $\Omega^{m}_{0} = \frac{8\pi G \rho^{m}_{0}}{3H^{2}_{0}}$, $\Omega^{X}_{0} = \frac{8\pi G \rho^{X}_{0}}{3H^{2}_{0}}$ and $\eta$ is a constant. Note that the continuity equation for anisotropy could be written as
\begin{equation}
\label{eq19}
\dot{\rho}^{\sigma}+6H\rho^{\sigma}=0,
\end{equation}
which in turn gives \cite{Amirhashchi/2014, Hassan/2018}
\begin{equation}
\label{eq20}
\rho^{\sigma}=\rho^{\sigma}_{0}a^{-6}.
\end{equation}
It is clear that the total energy density (combining the equations of motion for each species of density) satisfies the following differential equation:
\begin{widetext}
\begin{equation}
\label{eq21}
\frac{d\rho^{tot}}{da}+\frac{3}{a}\left[(1+\omega^{X})\frac{Aa^{\eta}}{1+Aa^{\eta}}(\rho^{tot}-\rho^{\sigma})+\frac{1}{1+Aa^{\eta}}(\rho^{tot}-\rho^{\sigma})+2\rho^{\sigma}\right]=0.
\end{equation}
\end{widetext}
After some algebra, we obtain

\begin{equation}
\label{eq22}
\frac{d\rho^{tot}}{da}+\frac{3}{a}\left[\frac{(1+\omega^{X})Aa^{\eta}\rho^{tot}+(1-\omega^{X})\rho^{\sigma}_{0}Aa^{\eta-6}}{1+Aa^{\eta}}\right]=0.
\end{equation}
Finally, the solution of above equation gives
\begin{align}
\label{eq23}
\begin{split}
\rho^{tot}&=\frac{1}{\eta-6}\biggl[(1+Aa^{\eta})^{\frac{-3(1+\omega^{X})}{\eta}}\biggl(3A\rho^{\sigma}_{0}a^{\eta-6}(\omega^{X}-1)\times\\&{}_2 F_1\biggl([\frac{\eta-6}{\eta},\frac{\eta-3(1+\omega^{X})}{\eta}],[\frac{2(\eta-3)}{\eta}], -Aa^{\eta}\biggr)\\&+C(\eta-6)\biggr)\biggr],
\end{split}
\end{align}
where $C$ is an integration constant and \textgravedbl ${}_2 F_1$\textacutedbl stands for hyper-geometric function. One can use equation (\ref{eq23}) in equations (\ref{eq18}) \& (\ref{eq13}) to obtain Hubble function which could be used in statistical analysis. However, in this case, the estimation of model parameters, numerically, will be very intensive. Therefore in what follows we study two special cases namely (1) the case in which the coupling is a constant, i.e., $\delta=\mbox{constant}$ and (2) the case in which $\delta$ is a function of time (redshift) and ratio of the DE-DM densities is $\rho^{m}/\rho^{X}\propto a^{-\eta}$.
\subsection{Constant coupling model}
\label{subsec:1}
For constant $\delta$, the first integral of equation (\ref{eq15}) reads as
\begin{equation}
\label{eq24}
\rho^{X} = \rho_{0}a^{\delta-3(1+\omega^{X})} = \rho_{0}(1+z)^{3(1+\omega^{X})-\delta},
\end{equation}
where $\rho_{0}$ is the constant of integration.\\
Equations (\ref{eq14}) and (\ref{eq24}) lead to
\[
\rho^{m} = \frac{\delta \rho_{0}}{3+\delta-3(1+\omega^{X})}a^{\delta-3(1+\omega^{X})}+\frac{k_{1}}{a^{3}}
\]
\begin{equation}
\label{eq25}
=\frac{\delta \rho_{0}}{3+\delta-3(1+\omega^{X})}(1+z)^{3(1+\omega^{X})-\delta}+k_{1}(1+z)^{3},
\end{equation}
where $k_{1}$ is also a constant of integration.\\
Thus, from Friedmann equation (\ref{eq13}), we obtain
\begin{align}
\begin{split}
\label{eq26}
H^{2} &= H_{0}^{2}\biggl[\Omega^{X}_{0}(1+z)^{3(1+\omega^{X})-\delta}+\\&
\frac{\delta \Omega^{(X)}_{0}}{\delta-3\omega^{(X)}}(1+z)^{3(1+\omega^{X})-\delta}+\\&
\left(\Omega^{m}_{0}-\frac{\delta \Omega^{X}_{0}}{\delta-3\omega^{X}}\right)(1+z)^{3}+\Omega^{(\sigma)}_{0}(1+z)^{6}\biggr],
\end{split}
\end{align}
where $\Omega^{(\sigma)}_{0}$ denotes the present value of energy density due to anisotropy of universe.\\
\subsection{Variable coupling model}
\label{subsec:2}
In this case, using the first integral of equation (\ref{eq15}), i.e., 
\begin{equation}
\label{eq27}
\rho^{X}=\rho_{0}a^{-3(1+\omega^{X})}\exp \left[-\int \delta d(ln a)\right],
\end{equation}
in equation (\ref{eq14}) and integrating we obtain

\begin{equation}
\label{eq28}
\rho^{m} = \rho^{m}_{0}a^{-3}\exp \left[\int \delta a^{\eta} d(ln a)\right],
\end{equation}
where we have also used definition of equation (\ref{eq17}). Note that from equations (\ref{eq27}) \& (\ref{eq28}), for constant $\omega^{X}$ and in the absence of coupling $\delta$, the energy densities of dark energy and dark matter scale as $\rho^{X}\propto a^{-3(1+\omega^{X})}$ \& $\rho^{m}\propto a^{-3}$ respectively. It is easy to conclude that $\rho^{X}/\rho^{m}=a^{-3\omega^{X}}$ is corresponding to $\eta=-3\omega^{X}$ in equation (\ref{eq17}). Therefore, it is clear that for an interacting scenario, the condition $\eta\neq-3\omega^{X}$ should be satisfied. Since $\Gamma=H\delta$, from equation (\ref{eq17}), we obtain 
\begin{equation}
\label{eq29}
\Gamma = -H(\eta+3\omega^{X})\Omega^{X}(z),
\end{equation}
where, from equation (\ref{eq17}),
\begin{equation}
\label{eq30}
\Omega^{(X)}(z) = \frac{1-\Omega^{\sigma}}{[\rho^{m}_{0}/\rho^{X}_{0}(1+z)^{\eta}+1]}. 
\end{equation}
Finally, from equations (\ref{eq29}) \& (\ref{eq30}), we obtain the coupling $\delta(z)$ as
\begin{equation}
\label{eq31}
\delta(z) = \frac{\delta_{0}}{[\Omega^{X}_{0}+(1-\Omega^{X}_{0}-\Omega^{\sigma}_{0})(1+z)^{\eta}]}, 
\end{equation}
where $\delta_{0}=-(\eta+3\omega^{X})(1-\Omega^{\sigma}_{0})\Omega^{X}_{0}$.\\
Now, we assume that the dark energy
and anisotropy are coupled with following relation:
\begin{equation}
\label{eq32}
\frac{\rho^{X}}{\rho^{\sigma}} = Ba^{\gamma}, ~~~B\equiv\frac{\rho^{X}_{0}}{\rho^{\sigma}_{0}}=\frac{\Omega^{X}_{0}}{\Omega^{\sigma}_{0}}. 
\end{equation}
Substituting equations (\ref{eq14}), (\ref{eq15}), (\ref{eq17}), (\ref{eq19}) and (\ref{eq32}), the total energy density, $\rho^{T}=\rho^{m}+\rho^{X}+\rho^{\sigma}$, satisfies
\begin{equation}
\label{eq33}
\frac{d\rho^{tot}}{\rho^{tot}} = -3\frac{da}{a}\left[1+\omega^{X}\left(1+2\frac{\rho^{\sigma}_{0}}{\rho^{X}_{0}}a^{-\gamma}+\frac{\rho^{m}_{0}}{\rho^{X}_{0}}a^{-\eta}\right)^{-1}\right]
\end{equation}

Hence, the expression for Hubble parameter for variable coupling model is obtained as
\begin{align}
\label{eq34}
\begin{split}
H^{2}& = H_{0}^{2}\exp\left[\int\frac{-3-3\omega^{X}\left(1+2\frac{\rho^{\sigma}_{0}}{\rho^{X}_{0}}a^{-\gamma}+\frac{\rho^{m}_{0}}{\rho^{X}_{0}}a^{-\eta}\right)^{-1}}{a\rho_{0}^{T}}da\right].
\end{split}
\end{align}
As above equation does not have an explicit analytical solution, we solve it numerically for the purpose of MCMC analysis. It is worth noting that since, without interaction, the energy densities of dark energy and anisotropy scale as $\rho^{X}\propto a^{-3(1+\omega^{X})}$ \& $\rho^{\sigma}\propto a^{-6}$ respectively, one can conclude $\gamma=3(1-\omega^{X})$. Also, since equation (\ref{eq34}) does not have an analytical solution, we use numerical solution in our MCMC code. To study the coupling affects on the evolution of some important cosmological quantities such as Hubble constant, $H_{0}$ (and deal with the tension problem of this parameter) and the dark energy equation of state parameter, $\omega^{X}$, in the next section, we place observational constraints on the strength of the coupling.
\section{DATA AND METHOD}
\label{data}
In this section, we shall briefly introduce the observational data and the statistical methodology to constrain the two
interacting scenarios discussed in previous section (we refer the reader to Ref. \cite{Amirhashchi/2019a} for details of likelihoods for OHD, BAO, CMB data and to Ref. \cite{Amirhashchi/2019b} for SNIa data).\\
\begin{itemize}

\item {\bf Observational Hubble Data (OHD)}: We adopt $31H(z)$ data points over the redshift range of $0.07\leq z\leq 1.965$ (see Table 2 of \cite{Ryan/2018}) obtained from cosmic chronometers (CC) technique. Since this technique is based on the \textquotedblleft galaxy differential age\textquotedblright method, OHD data obtained from CC technique is model-independent and all data of this compilation are uncorrelated.\\

\item {\bf Type Ia Supernovae (SNIa)}: We use the Pantheon compilation \cite{Scolnic/2018} including 1048 SNIa apparent magnitude measurements in the redshift range $0.01 < z < 2.3$ , which includes 276 SNIa $(0.03 < z < 0.65)$ discovered by the Pan-STARRS1 Medium Deep Survey and SNIa distance estimates from SDSS, SNLS and low-zHST samples.\\

\item {\bf Baryon acoustic oscillations (BAO)}: We consider 10 BAO data extracted from the 6dFGS \cite{Beutler/2012}, SDSS-MGS \cite{Ross/2015}, BOSS \cite{Alam/2017}, BOSS CMASS \cite{Anderson/2014}, and WiggleZ \cite{Kazin/2014} surveys.\\

\item {\bf Cosmic Microwave Background (CMB)}: While the CMB distance priors have been widely used to obtain cosmological constraints, it has recently been shown that for extended models these do not provide a good approximation for the full CMB likelihood \cite{Zhai/2019}. In this paper, we consider the latest high (and low)-$l$ temperature and polarization cosmic microwave background measurements from Planck \cite{Aghanim/2018a, Aghanim/2018b, Aghanim/2018c}. These data are very powerful to analyze the cosmological models with spatial anisotropy. \\

\item {\bf Hubble Space Telescope (HST)}: We also use the recent estimation of the Hubble constant, $H_{0}=74.02\pm1.42$ at $68\%$ confidence  level (CL) obtained from Hubble Space Telescope \cite{Riess/2019}. In this paper we refer to this data as R19. It is worth mentioning that this new estimation of $H_{0}$ is in tension with Planck's estimation within the minimal
$\Lambda$CDM model at $4.4\sigma$. 
\end{itemize}
For the statistical analyses, we use Pymc3 python package to generate MCMC chains using Metropolis-Hastings algorithm. The parameter space for the first case (constant coupling) is 
\begin{equation}
\label{eq35}
{\bf\Theta_{1}}= \{H_{0}, \Omega^{m}, \Omega^{X}, \omega^{X}, \delta\},
\end{equation}
and for the second case (variable coupling) is
\begin{equation}
\label{eq36}
{\bf\Theta_{2}}= \{H_{0}, \Omega^{m}, \Omega^{X}, \omega^{X}, \sigma_{0}, \eta, \gamma\}.
\end{equation}
Note that in both cases the anisotropy density, $\Omega^{\sigma}$, is a derived parameter. To stabilize our estimations, we run 4 parallel chains with 10000 iterations for each parameter. Moreover, we perform both well-known Gelman-Rubin and Geweke tests to confirm the convergence of the generated MCMC chains. For this purpose, we also monitor the trace plots for good mixing of the posterior distributions. We use python package GetDist \cite{getdist} for analysing MCMC chains.\\

Moreover, we perform co-variance matrix (which could be obtained from our MCMC runs) in order to check the degeneracy direction between computed parameters. Theoretically, co-variance matrix of the parameter space $\{\theta\}$ could be defined as: 
\begin{equation}
\label{eq37} C_{\alpha\beta}=\rho_{\alpha\beta}\sigma({\theta_{\alpha}})\sigma({\theta_{\beta}}),
\end{equation}
where $\sigma({\theta_{\alpha}})$ and $\sigma({\theta_{\beta}})$ give the uncertainties in parameters $\theta_{\alpha}$ and $\theta_{\beta}$ at $1\sigma$ error respectively, and $\rho_{\alpha\beta}$ is the correlation coefficient between $\theta_{\alpha}$ and $\theta_{\beta}$. Note that $\rho$ varies from 0 (independent) to 1 (completely correlated).\\
Finally, we consider the following uniform priors imposed on the model parameters in our statistical analyses.
\begin{table}[h!]
\caption{Uniform priors imposed on free	parameters of the interacting scenarios.}
\centering
\scalebox{0.8}{
\begin{tabular} {c|c}
\hline
Parameter&  Prior\\[0.5ex]
\hline
$H_{0}$&$[50,100]$\\
$\Omega^{m}$&$[0,1]$\\
$\Omega^{X}$&$[0,1]$\\
$\omega^{X}$&$[-2,0]$\\
$\delta, \delta_{0}$&$[-1,1]$\\
$\eta$&$[0,10]$\\
$\gamma$&$[0,10]$\\
[0.5ex]
\hline
\end{tabular}}
\label{tab:1}
\end{table}
\begin{figure*}[ht!]
\centering
\begin{subfigure}{0.3\textwidth}
\centering
\includegraphics[width=\textwidth]{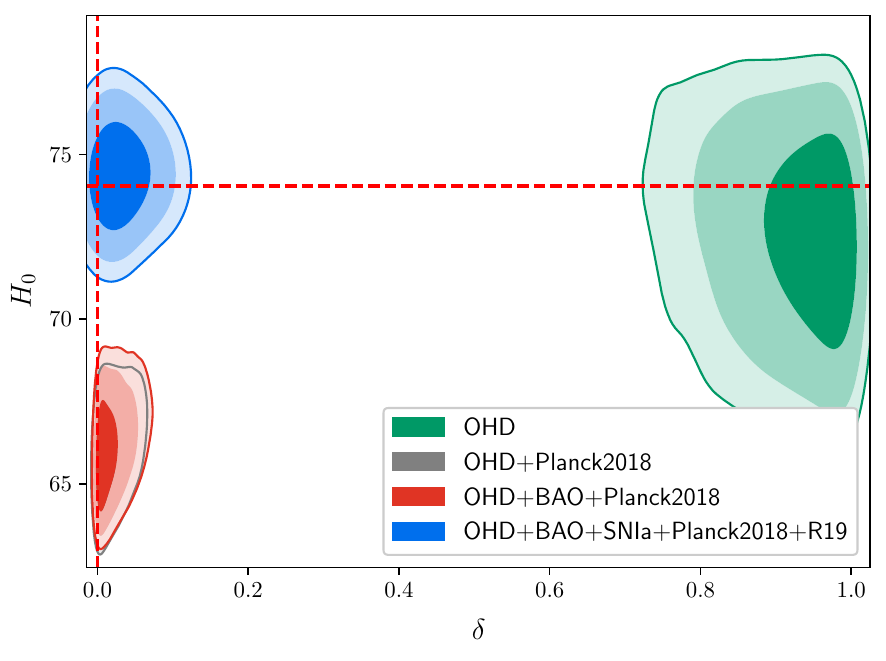}
\caption{\label{fig1a}}
		
\end{subfigure}%
\begin{subfigure}{0.3\textwidth}
\centering
\includegraphics[width=\textwidth]{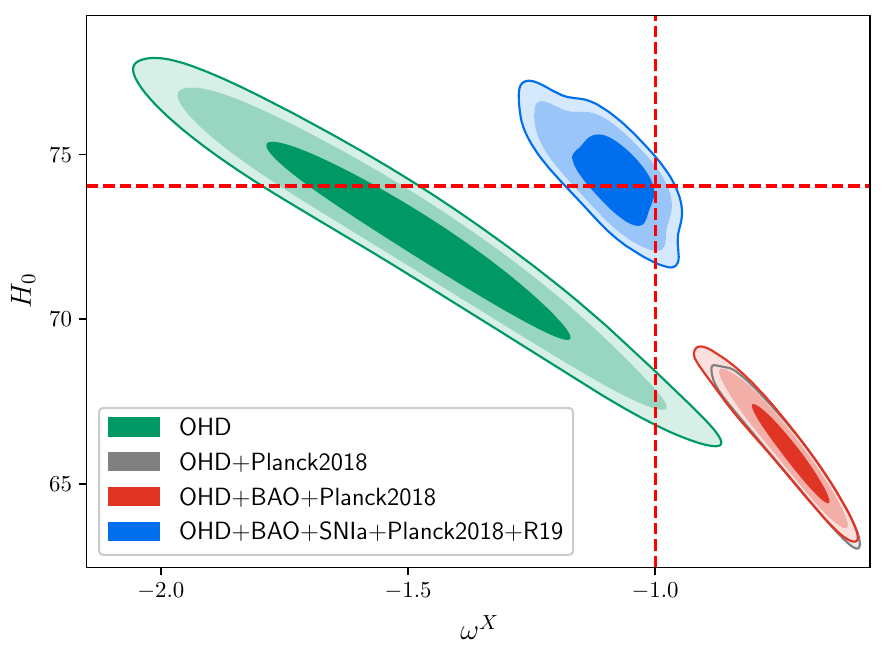}
\caption{\label{fig1b}}
		
\end{subfigure}
\begin{subfigure}{0.3\textwidth}
\centering
\includegraphics[width=\textwidth]{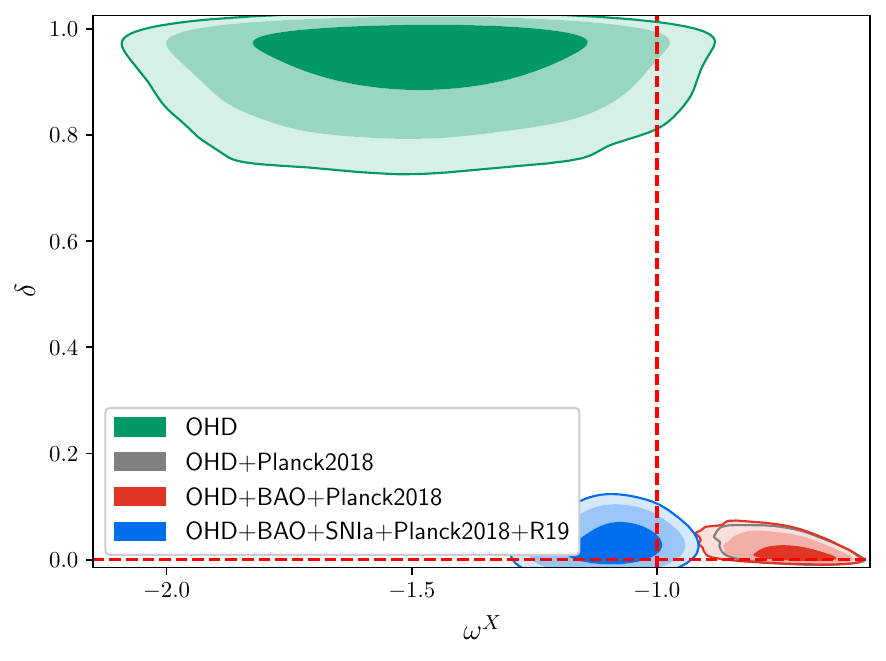}
\caption{\label{fig1c}}
		
\end{subfigure}
\caption{\label{fig1} The contour plots of (a) $H_{0}-\delta$ plane, (b) $H_{0}-\omega^{X}$ plane, and (c) $\delta-\omega^{X}$ plane at $1\sigma$-$3\sigma$ confidence levels. The dashed horizontal and vertical lines in (a)\& (b) show the mean value of Hubble constant obtained from HST project and the mean values of $\delta$ \& $\omega^{X}$ obtained in this work respectively. In (c), the mean estimated values of $\delta$ and $\omega^{X}$ are shown by horizontal and vertical lines respectively.}	
\end{figure*}
\section{Results}
\label{results}
In this section, we use the observational data presented in the previous section to study the viability of both theoretical interacting models. Specially we focus on the study (constraints) of three important parameters namely coupling, DE equation of state, and Hubble constant parameters. We also deal with the problem of Hubble constant tension in our study. On comparing the result obtained in this paper with the existing isotropic models with interacting dark sectors (Nunes \& Valentino \cite{Nunes/2021R} and Valentino et al. \cite{Valentino/2020R}), we observe that in Nunes \& Valentino \cite{Nunes/2021R}, $H_{0}$ is constrained as $H_{0} = 68.5^{+1.2}_{-1.1}$ with $1\sigma$ error for the Pantheon+BAO+BBN joint analysis while in Valentino et al. \cite{Valentino/2020R}, $H_{0} = 71.1 \pm 1.1$ for the Planck + BAO + R19 data. In this paper, we have obtained $H_{0} = 73.9 ^{+ 1.5}_{- 0.95}$ and $H_{0} = 69.73 \pm {0.67}$ for joint OHD+CMB+BAO+SNIa+R19 analysis for both constant and varying coupling models respectively. Further, we observe that the value of $H_{0}$ in constant coupling model finds 3.2 $\sigma$ tension and 0.92 $\sigma$ tension with its corresponding values obtained in Nunes \& Valentino \cite{Nunes/2021R} and Valentino et al. \cite{Valentino/2020R} respectively. For varying coupling model, the tensions in $H_{0}$ value are quantified as 1.7 $\sigma$ and 1.06 $\sigma$ from its corresponding values in isotropic $\Lambda$CDM model \cite{Nunes/2021R,Valentino/2020R}. It has been also observed that in Nunes \& Valentino \cite{Nunes/2021R}, the coupling parameter is constrained to be consistent with zero. Here, we have constrained coupling parameter as $\delta = 0.023^{+0.065}_{-0.034}$ and $\delta_{0} = 0.013^{+0.040}_{-0.055}$ at $1\sigma$ error for both the constant and variable coupling model respectively. Recently, Kumar \cite{Kumar/2021} has also explored some features and consequences of a phenomenological interaction in dark sector for isotropic universe and constrained $H_{0} = 72.8^{+1.4 + 2.8}_{-1.4 - 2.7}$ and coupling parameter $\xi = -0.41^{+0.12 + 0.22}_{-0.12 - 0.22}$ with Planck+R19 data. The values of $H_{0}$ obtained in this paper for constant and variable coupling model are in $1.97 ~\sigma$ and $0.59~ \sigma$ tensions with the estimated value of $H_{0}$ in Kumar \cite{Kumar/2021}. It is worthwhile to note that  there is the massive shift in preferred parameters when only OHD are used. The reason is that OHD correspond to low redshifts and are therefore unable to put tight constraints on the model parameters. On the other hand, CMB and BAO data likelihoods include fixed high redshifts such as the drag redshift and the last scattering redshift and, therefore, preserve the standard $\Lambda$CDM evolution of the universe at early times leading to tight constraints on the model parameters (see \cite{ozgur19}). Therefore, only the OHD data can not be used as a complete package for constraining the various parameters of anisotropic universe. The detailed results of observational analyses for constant and varying coupling models are described below.
\subsection{Constant Coupling Model}
\label{const} 
In Table~\ref{tab:2}, we have listed the results of our statistical analysis (observational constraints) on the constant coupling model for different data set and their joint combination at $1\sigma$ CL. Let us first discuss the observational constraints on the dark coupling parameter $\delta$.
\begin{table*}[htb!]
\caption{Best fit values for the constant coupling model parameters at 1$\sigma$ error bars.}
\centering
\centering
\scalebox{0.7}{
\begin{tabular} {ccccccccc}
\hline
Parameter    & OHD & OHD+CMB & OHD+BAO &OHD+CMB+BAO & OHD+CMB+BAO+SNIa&OHD+CMB+BAO+R19 & OHD+CMB+BAO+SNIa+R19\\[0.5ex]           
\hline
\hline{\smallskip}
$H_{0}$ & $72.5^{+2.1}_{-1.9}$ & $65.82^{+0.85}_{-0.99}$&$64.20^{+0.92}_{-1.1}$ & $66.01^{+0.96}_{-1.1}$& $70.4^{+2.1}_{-2.6}$ & $70.6^{+1.5}_{-2.0}$&$73.9^{+1.5}_{-0.95}$\\[0.3cm]              
			
$\Omega^{m}$ &  $0.3017\pm 0.0057$  & $0.3^{+0.00021}_{-0.00023}$&$0.3001\pm 0.0029$ & $0.3\pm0.0048$ & $0.294^{+0.057}_{-0.042}$ &$0.291\pm 0.025$ & $0.282^{+0.065}_{-0.034}$\\[0.3cm]        
			
$\delta$ &  $0.941^{+0.059}_{-0.012}$ & $0.0134^{+0.0023}_{-0.013}$&$0.0265^{+0.0062}_{-0.026}$ & $0.0140^{+0.0026}_{-0.014}$&$0.046\pm0.043$ & $0.0368^{+0.0072}_{-0.040}$& $0.023^{+0.017}_{-0.024}$\\[0.3cm]  
			
$\Omega^{X}$ & $0.7039\pm 0.0056$ & $0.7^{+0.00034}_{-0.00025}$&$0.6999^{+0.0032}_{-0.0029}$  & $0.7^{+0.001}_{-0.0012}$&$0.706^{+0.042}_{-0.057}$ & $0.709\pm 0.025$& $0.717^{+0.034}_{-0.065}$\\[0.3cm]
			
$\omega^{X}$ & $-1.47^{+0.21}_{-0.19}$ & $-0.720^{+0.055}_{-0.040}$&$-0.650^{+0.057}_{-0.042}$ & $-0.730^{+0.060}_{-0.047}$&$-0.995^{+0.056}_{-0.067}$ & $-0.99\pm 0.10$& $-1.105\pm 0.075$\\[0.3cm]
			
$\Omega^{\sigma}$ & $-0.00569^{+0.00029}_{-0.00034}$ & $\left(\,1.0^{+3.2}_{-1.1}\,\right)\cdot 10^{-10}$&$0.000088^{+0.000020}_{-0.000088}$ & $\left(\,3.6^{-2.3}_{-3.6}\,\right)\cdot 10^{-10}$&$\left(\,2.4^{-1.2}_{-2.5}\,\right)\cdot 10^{-7}$ & $\left(\,1.7^{+1.2}_{-1.9}\,\right)\cdot 10^{-7}$& $\left(\,4.73^{+0.75}_{-4.4}\,\right)\cdot 10^{-5}$\\[0.5ex]
			
\hline
\end{tabular}}
\label{tab:2}
\end{table*}
From Table~\ref{tab:2}, we observe that using OHD data alone results in very high value of coupling constant, i.e., $\delta=0.941^{+0.059}_{-0.012}$. It could be seen from this table that adding CMB (Planck 2018) or BAO to OHD data significantly decreases the coupling parameter $\delta$. Also our computations show that inclusion of BAO data to the OHD+CMB data does not affect considerably the estimated value of coupling constant. When we include Gaussian prior on the Hubble constant (R19)\cite{Riess/2019} in to our MCMC computation code, the joint combination of OHD+BAO+CMB data estimation gives $\delta=0.0368^{+0.0072}_{-0.040}$. Finally, from the joint analysis with OHD+BAO+SNIa+CMB+R19, we find $\delta=0.046\pm0.043$($0.023^{+0.017}_{-0.024}$), which also does not lead to any evidence for $\delta\neq 0$ and hence any possible interaction between dark components.\\
\begin{figure*}[ht!]
\centering
\begin{subfigure}{0.3\textwidth}
\centering
\includegraphics[width=\textwidth]{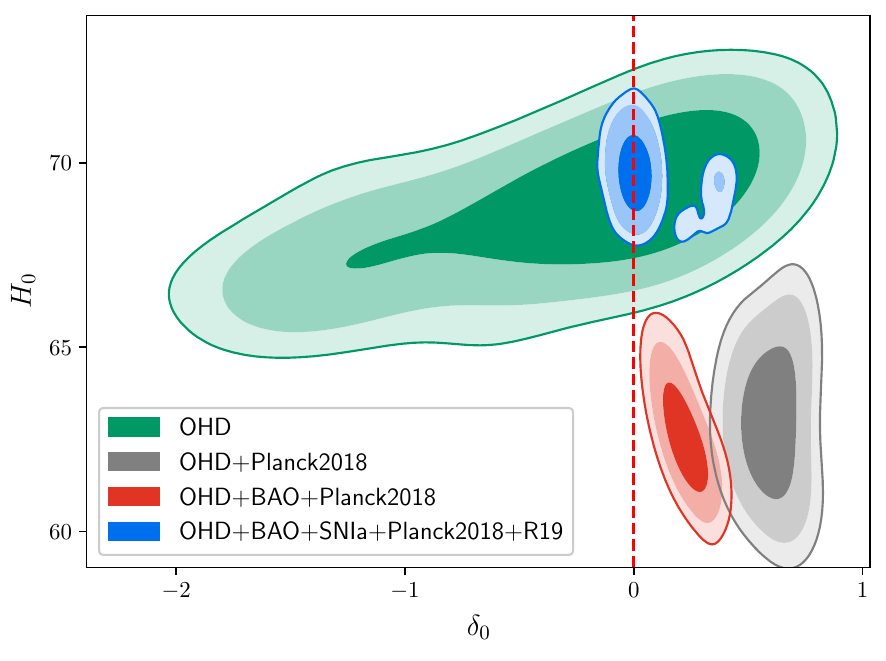}
\caption{\label{fig4a}}
		
\end{subfigure}%
\begin{subfigure}{0.3\textwidth}
\centering
\includegraphics[width=\textwidth]{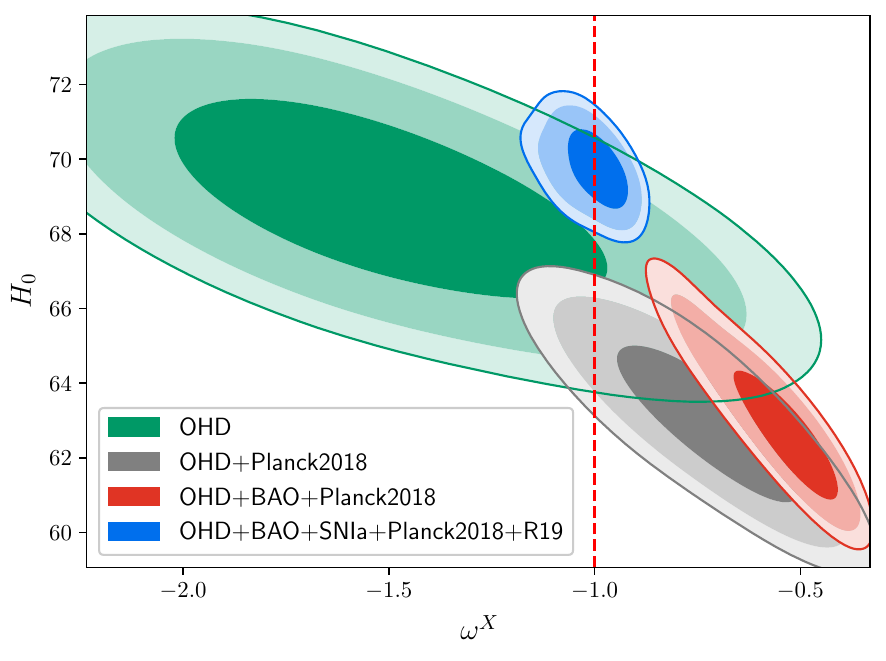}
\caption{\label{fig4b}}
		
\end{subfigure}
\begin{subfigure}{0.3\textwidth}
\centering
\includegraphics[width=\textwidth]{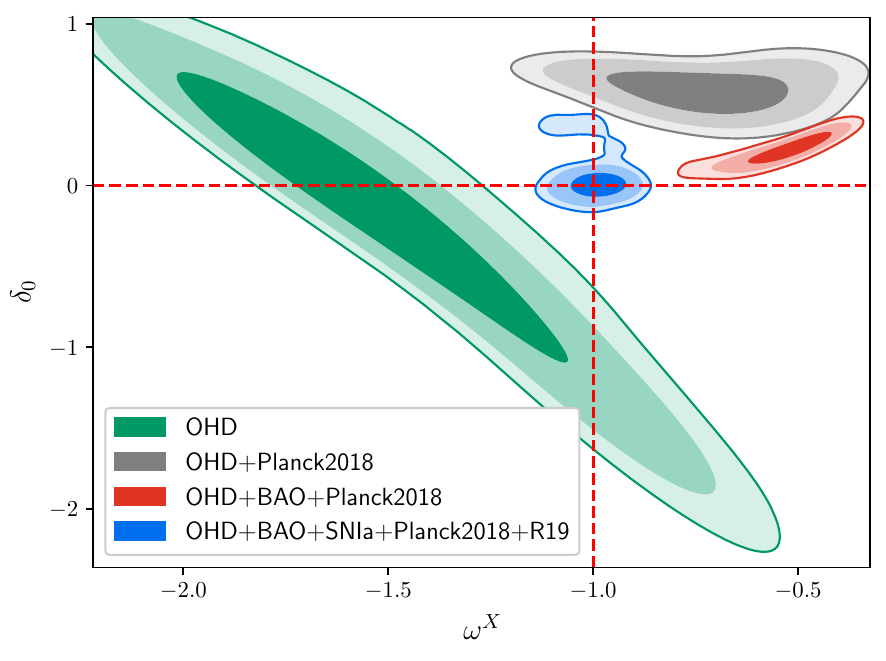}
\caption{\label{fig4c}}
		
\end{subfigure}
	
\begin{subfigure}{0.3\textwidth}
\centering
\includegraphics[width=\textwidth]{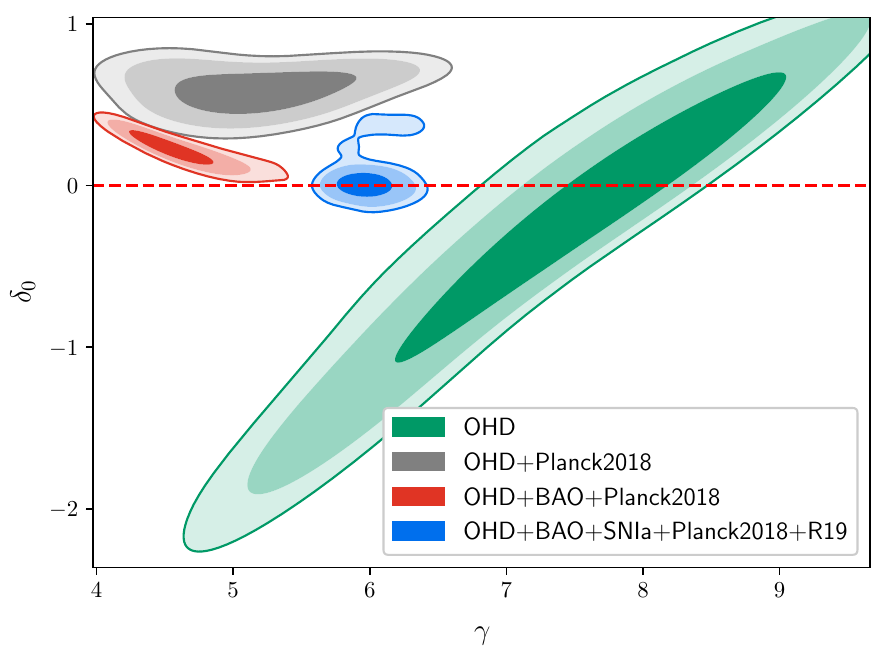}
\caption{\label{fig4d}}
		
\end{subfigure}
\caption{\label{fig4} The contour plots of (a) $H_{0}-\delta_{0}$ plane, (b) $H_{0}-\omega^{X}$ plane, (c) $\delta_{0}-\omega^{X}$ and (d) $\delta_{0}-\gamma$ planes at $1\sigma$-$3\sigma$ confidence levels. The dashed vertical, in figure (a) , and horizontal, in figures (c) and (d), lines stand for $\delta_{0}=0$. In (b) and (c), the dashed vertical lines stand for $\omega^{X}=-1$.}	
\end{figure*}
It is worth to mention that, it has already been argued by Guo et al. \cite{Guo/2007} that while including data in a high-redshift region ($z\gg1$) rules out models with strong couplings, BAO data do not provide stringent constraints on $\delta$. Nevertheless, our estimations show that BAO data also put tight constraints on coupling constant. The reason for different results with BAO data in Guo et al. \cite{Guo/2007} could be the usage of old BAO data with an optimization method rather than MCMC one. Figure~\ref{fig1a}, \ref{fig1a} \& \ref{fig1c} show the $1\sigma-3\sigma$ contour plots of ($H_{0}-\delta$), ($H_{0}-\omega^{X}$) and ($\delta-\omega^{X}$) pairs for different data sets, respectively. We notice positive mean values of $\delta$ in our results, indicating the energy/momentum transfer from the dark energy to dark matter. It has recently been argued by some authors (for example see \cite{Kumar/2019, Poulin/2019, Khosravi/2019, Yang/2019a, Yang/2019b, Arenas/2018, Vattis/2019, Yang/2018}) that a possible interaction between dark sectors can solve the current observational Hubble tension present in the $\Lambda$CDM model. In fact, these researches show that there is a positive correlation between $H_{0}$ and $\delta$ which means that higher values of Hubble constant require higher values of coupling parameter as well. The results of these studies also show that there is a negative correlation between $H_{0}$ and DM density $\Omega^{m}$. For instance, in Ref. \cite{Kumar/2019}, from joint combination of Planck+HST+KiDS, it is obtained $H_{0}=73.6^{+1.6}_{-1.6}$, $\delta=-0.40^{+0.16}_{-0.14}$ \& $\Omega^{m}=0.262^{+0.010}_{-0.012}$. From Table~\ref{tab:2}, it is observed that while the computed $H_{0}$ for OHD data alone is in good agreement with recent local measurement from HST but it requires a high value of the coupling parameter $\delta$. Using joint combination of all data plus R19 results in a value of $H_{0}$, with negligible coupling constant, which is in excellent agreement with what is reported by Riess et al (HST) \cite{Riess/2019}. The constraints on the Hubble constant for all combinations of data sets considered in this work are shown in Figure~\ref{fig1}. As it could be seen from Table~\ref{tab:2} and Figures~\ref{fig1a} \& \ref{fig1b}, there is no significant correlation between $H_{0}$ and $\delta$. However, except for joint combination of all data sets + R19, there is a significant correlation between $H_{0}$ and $\omega^{X}$. From Figure~\ref{fig1b} and also Table~\ref{tab:2}, it is clear to see that, except for joint data set + R19, a higher value of $H_{0}$ requires a lower value of EOS parameter $\omega^{X}$. In fact any DE model could be characterized by it's EOS parameter. Considering $\omega^{X}=-1$ as cosmological constant model (sometimes refer to this as phantom divided line-PDL), any model with $-\frac{1}{3}>\omega^{X}>-1$ refers to quintessence \cite{Wetterich/1998, Ratra/1988} and models with $\omega^{X}<-1$ are called phantom dark energy models \cite{Caldwell/2002}. It is important to note that as shown by Carroll et al \cite{Carroll/2003}, phantom fields are generally plagued by ultraviolet quantum instabilities. There is also DE scenario with EOS parameter varying between quintessence and phantom regions dubbed Quintom \cite{Feng/2005}. Our estimations show that for OHD alone, we have $-1.66<\omega^{X}<-1.26$, which means that DE completely varies in the phantom region. For OHD+BAO+SNIa+CMB+R19 data, EOS parameter, at $68\%$ CL, vary in interval $-1.125<\omega^{X}<-1.03$, which is in good agreement with both 9 years WMAP \cite{Hinshaw/2013} and Planck 2018 \cite{Aghanim/2018a} results. For all other data sets, the estimated EOS parameter is $\omega^{X}\sim -1$, which show that our DE model represents almost cosmological constant scenario. Figure~\ref{fig3} depicts the correlation matrix for combination of different data sets. From this figure, it is obvious that there is a more or less meaningful correlation between $\Omega^{m}$ and $H_{0}$ for joint combination of all data (with and without R19). However, this correlation is low enough to avoid the discrepancy of matter density mentioned in previous literature such as \cite{Kumar/2019, Poulin/2019, Khosravi/2019, Yang/2019a, Yang/2019b}. For example, as mentioned above, in \cite{Kumar/2019}, the estimated values for Hubble and DM density parameters are $H_{0}=73.6^{+1.6}_{-1.6}$ \& $\Omega^{m}=0.262^{+0.010}_{-0.012}$ respectively, which means that a High value of $H_{0}$ requires a low value of $\Omega^{m}$. \\
\begin{figure*}[ht!]
\centering
\begin{tabular}{@{}cccc@{}}
\includegraphics[width=.3\textwidth]{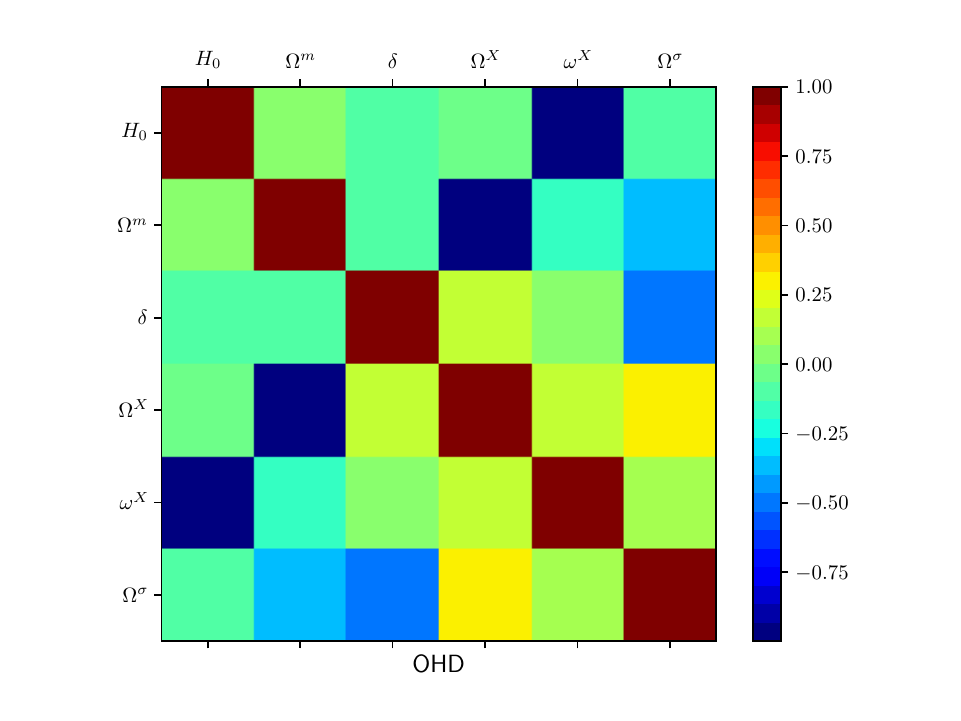} &
\includegraphics[width=.3\textwidth]{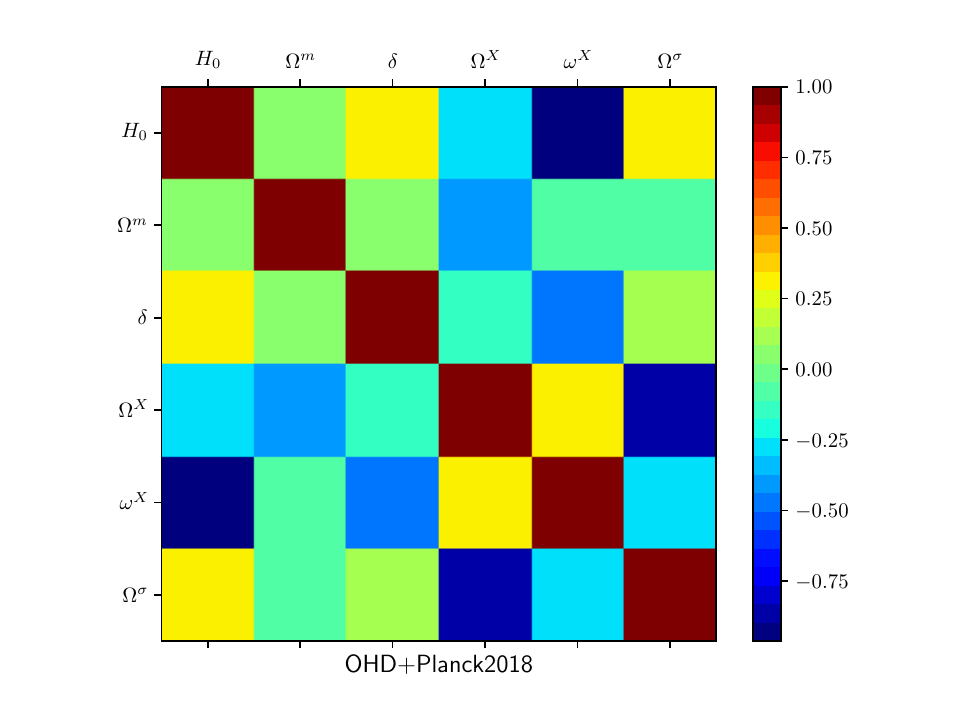} &
\includegraphics[width=.3\textwidth]{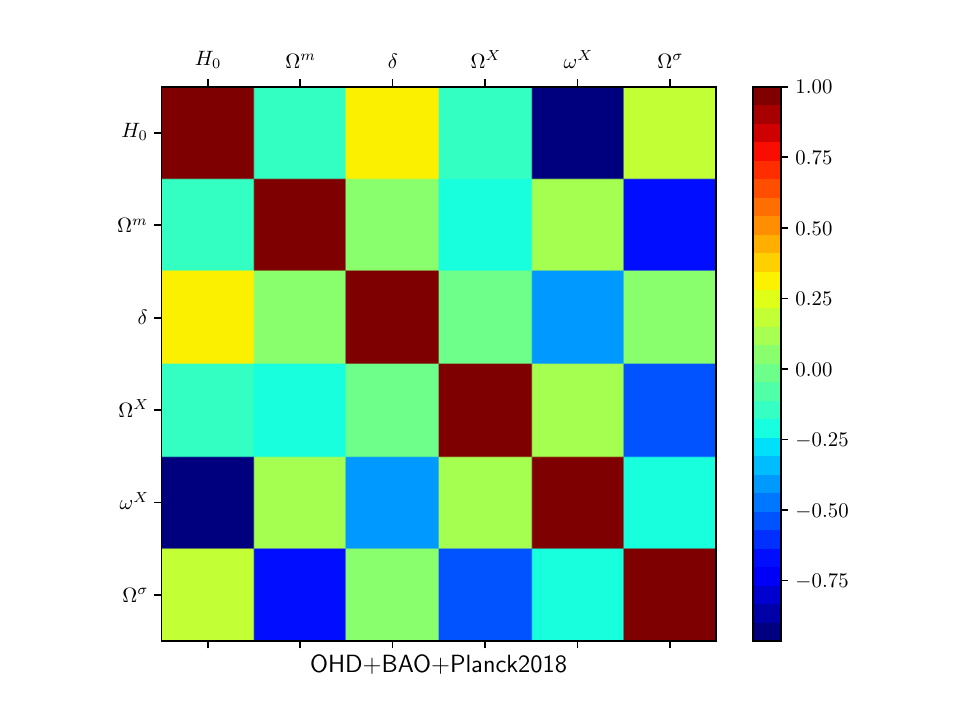} \\
\includegraphics[width=.3\textwidth]{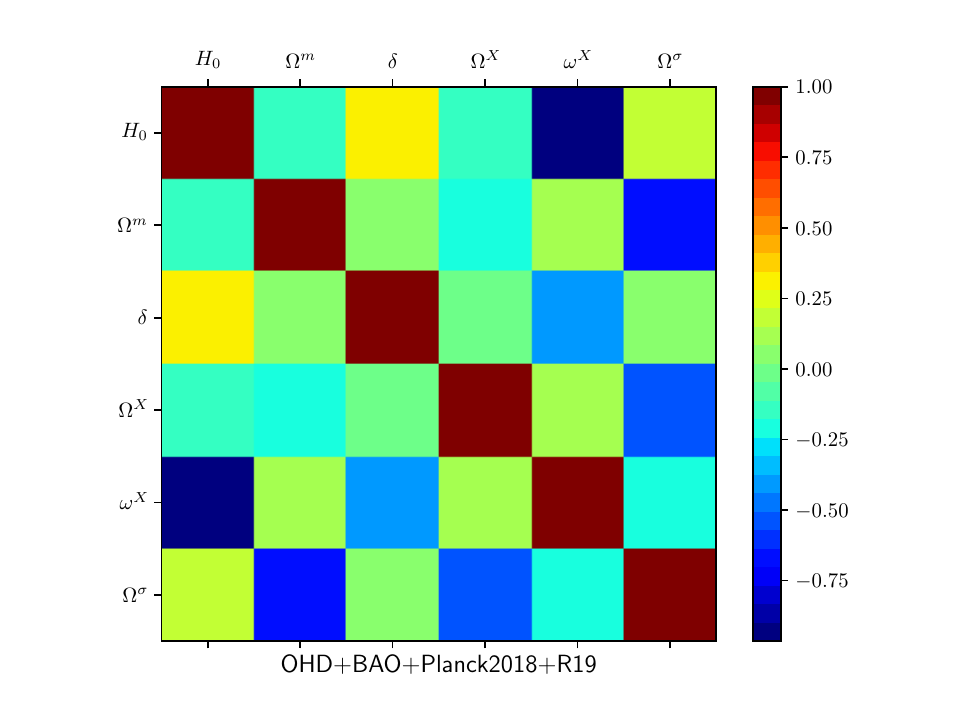} &
\includegraphics[width=.3\textwidth]{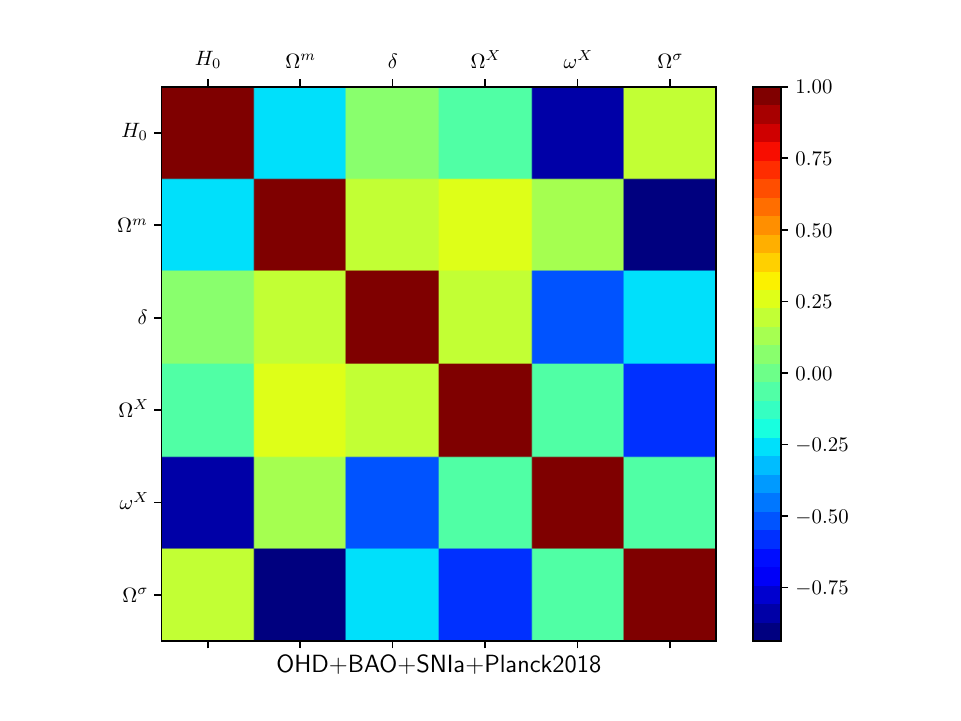} &
\includegraphics[width=.3\textwidth]{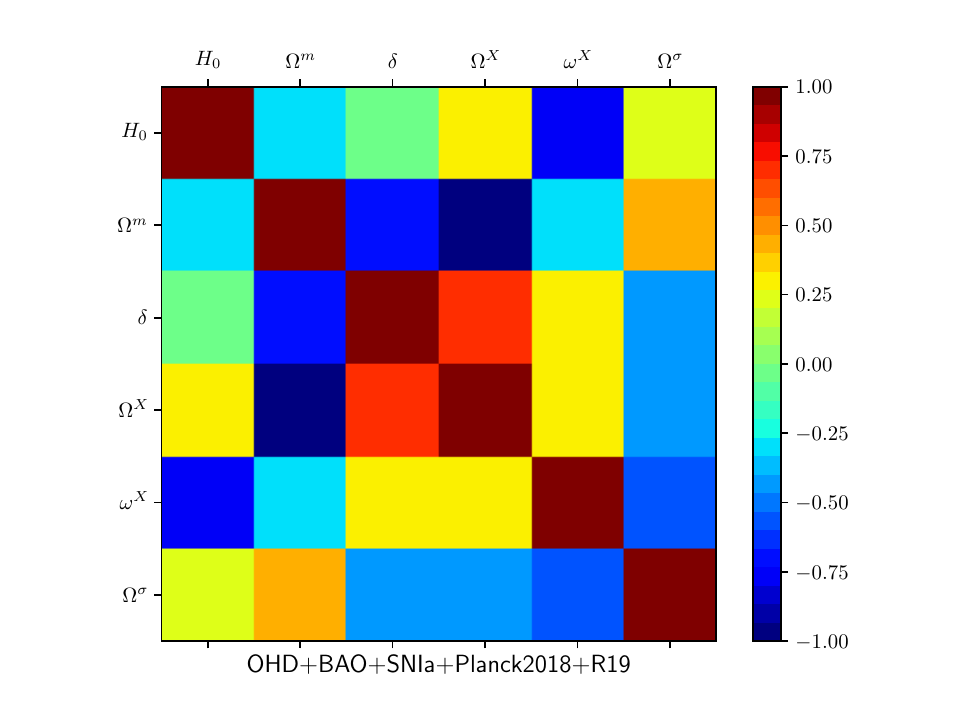}\\
\end{tabular}
\caption{Plots of correlation matrix of parameter space ${\bf\Theta_{1}}$ using combinations of different data for constant coupling model.	The color bars share the same scale.}
\label{fig3}
\end{figure*}

\subsection{Varying Coupling Model}
\label{vary}
\begin{figure}[h!]
\includegraphics[width=9cm,height=8cm,angle=0]{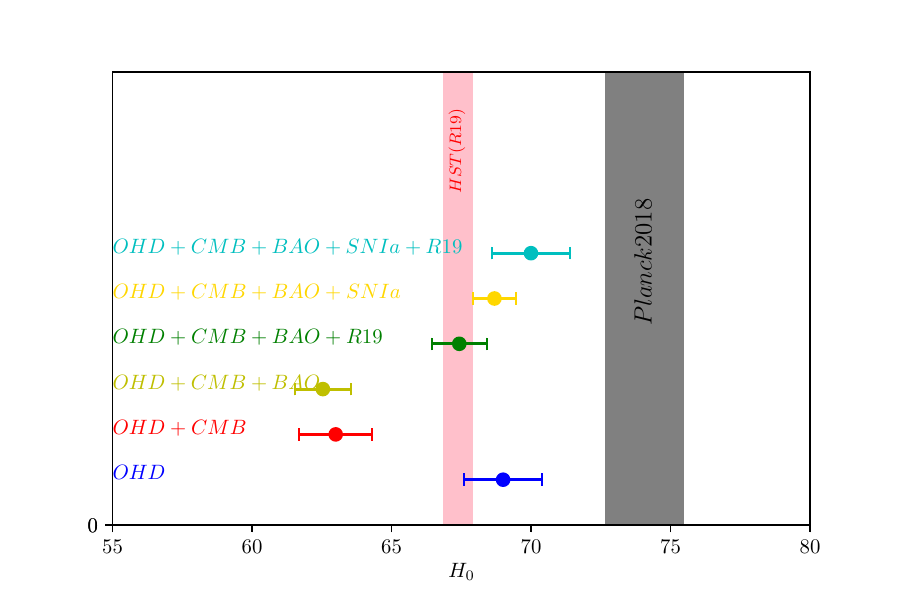}
\caption{Whisker plot with the $1\sigma$ confidence  level on the Hubble constant for varying coupling case. Pink and gray vertical bands correspond to the value for the Hubble constant estimate by the Planck 2018 release \cite{Aghanim/2018a}and the HST in \cite{Riess/2019} respectively.}
\label{fig5}
\end{figure}
\begin{figure*}[ht!]
\centering
\begin{tabular}{@{}cccc@{}}
\includegraphics[width=.3\textwidth]{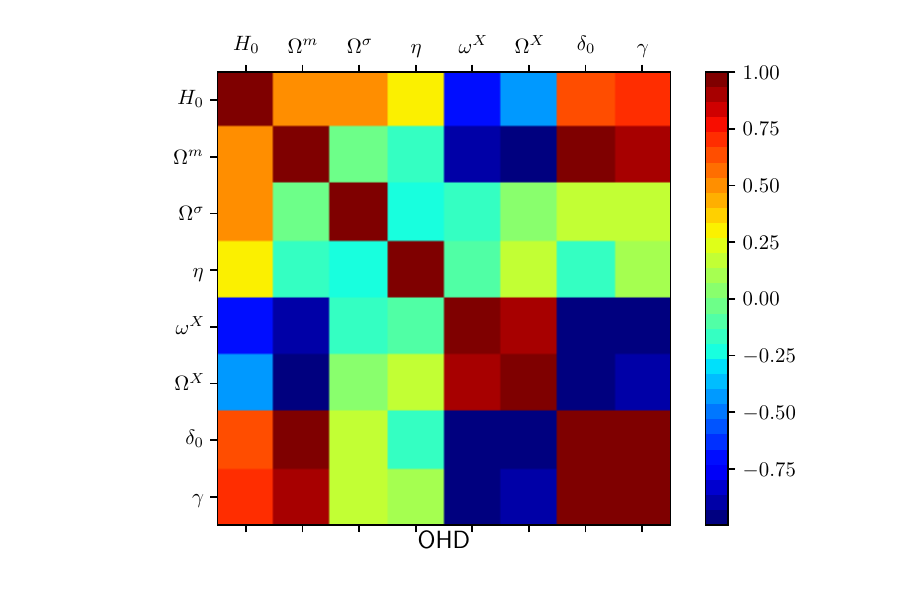} &
\includegraphics[width=.3\textwidth]{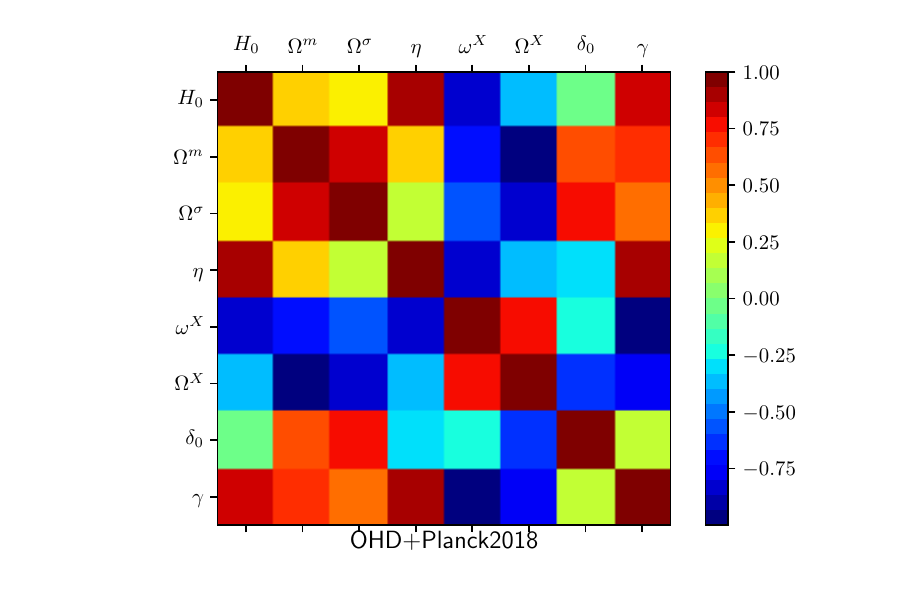} &
\includegraphics[width=.3\textwidth]{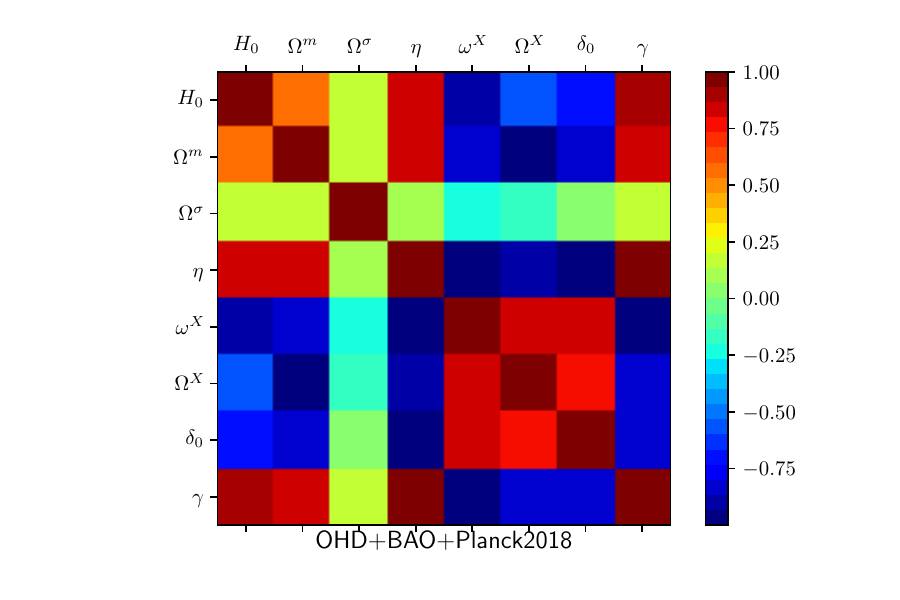} \\
\includegraphics[width=.3\textwidth]{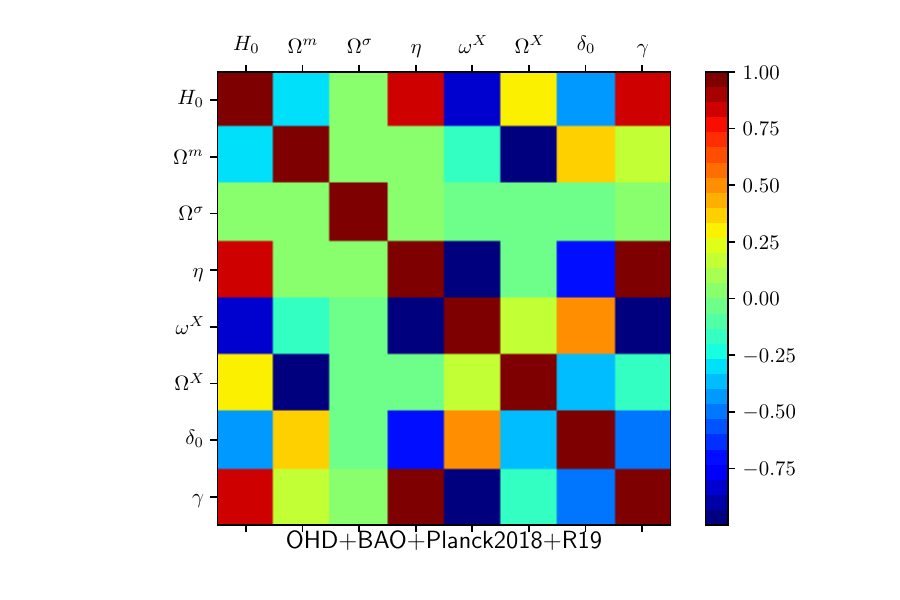} &
\includegraphics[width=.3\textwidth]{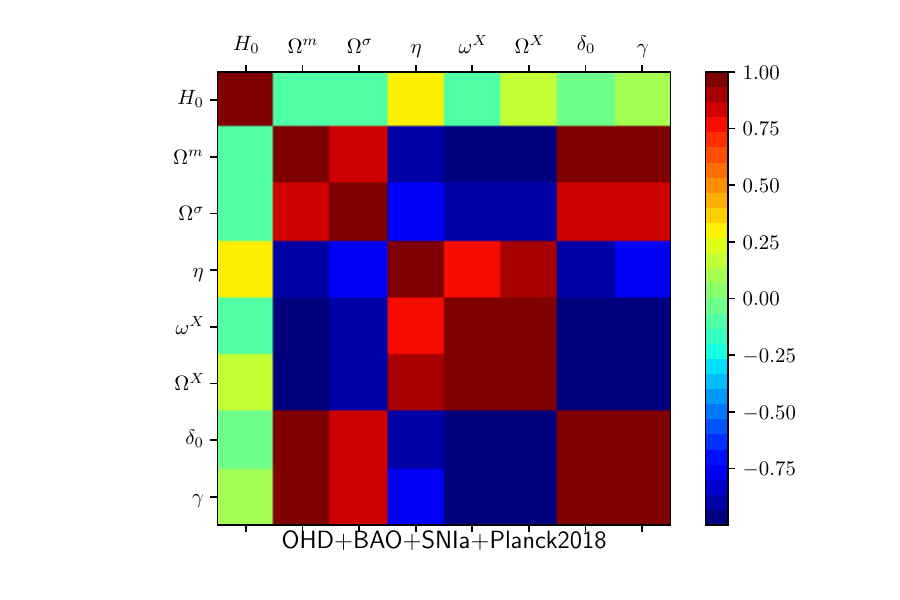} &
\includegraphics[width=.3\textwidth]{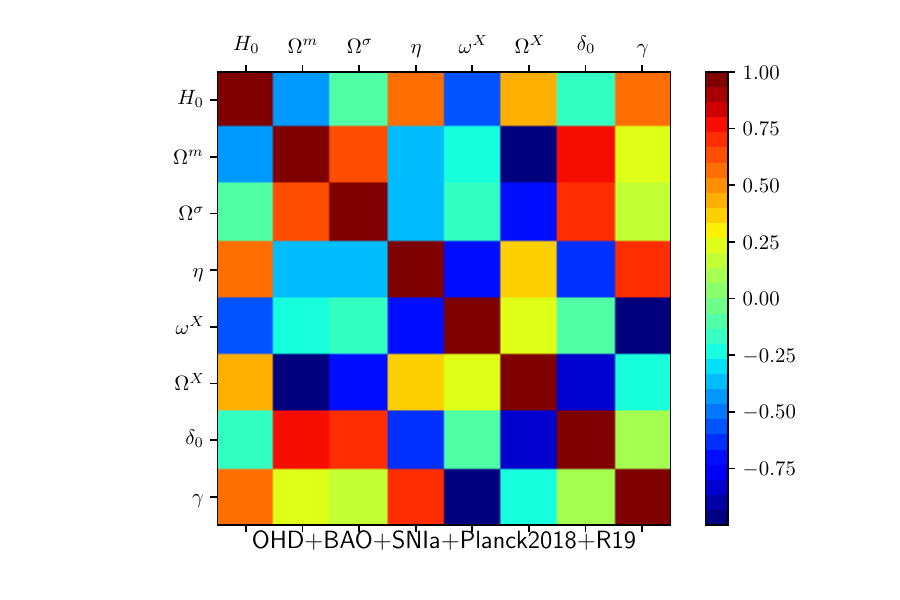}\\
\end{tabular}
\caption{Plots of correlation matrix of parameter space ${\bf\Theta_{1}}$ using combinations of different data for varying coupling model.	The color bars share the same scale.}
\label{fig6}
\end{figure*}
\begin{table*}[htb!]
\caption{Best fit values for varying coupling model parameters at 1$\sigma$ error bars.}
\centering
\centering
\scalebox{0.7}{
\begin{tabular} {ccccccccc}
\hline
Parameter   & OHD & OHD+CMB & OHD+BAO &OHD+CMB+BAO & OHD+CMB+BAO+SNIa&OHD+CMB+BAO+R19 & OHD+CMB+BAO+SNIa+R19\\[0.5ex]           
\hline
\hline{\smallskip}
$H_{0}$ & $68.9\pm 1.4$ & $63.0\pm 1.3$&$62.55\pm 0.97$ & $62.54^{+0.81}_{-1.0}$& $68.69\pm 0.76$ & $67.43\pm 0.98$&$69.73\pm 0.67$\\[0.3cm]              
			
$\Omega^{m}$ & $0.441^{+0.083}_{-0.030}$  & $0.464\pm 0.040$&$0.51\pm 0.11$ & $0.299^{+0.026}_{-0.022}$ & $0.428\pm 0.090$ &$0.3181\pm 0.0098$ & $0.3142^{+0.0076}_{-0.012}$\\[0.3cm]        
			
$\Omega^{\sigma}$ & $-0.00646\pm 0.00055$ & $\left(\,2.7^{+1.3}_{-2.1}\,\right)\cdot 10^{-5}$&$0.000142^{+0.000038}_{-0.00015}$ & $\left(\,0.1\pm 2.4\,\right)\cdot 10^{-8}$&$\left(\,4.7\pm 5.2\,\right)\cdot 10^{-5}$ & $\left(\,1.39^{+0.10}_{-1.1}\,\right)\cdot 10^{-11}$& $\left(\,0.5\pm 3.2\,\right)\cdot 10^{-7}$\\[0.3cm] 

$\eta$ &  $4.78^{+0.22}_{-0.055}$ & $1.09^{+0.41}_{-0.34}$&$0.89\pm 0.33$ & $1.26^{+0.25}_{-0.31}$&$2.52\pm 0.31$ & $2.40\pm 0.24$& $2.95\pm 0.17$\\[0.3cm] 
			
$\omega^{X}$ & $-1.49^{+0.19}_{-0.39}$ & $-0.74^{+0.16}_{-0.14}$&$-0.79^{+0.24}_{-0.16}$  & $-0.526^{+0.077}_{-0.058}$&$-1.18\pm 0.20$ & $-0.829\pm 0.067$& $-0.990^{+0.047}_{-0.040}$\\[0.3cm]
			
$\Omega^{X}$ & $0.565^{+0.031}_{-0.083}$ & $0.536\pm 0.040$&$0.49\pm 0.11$ & $0.701^{+0.022}_{-0.026}$&$0.572\pm 0.090$ & $0.6819\pm 0.0098$& $0.686^{+0.012}_{-0.0076}$\\[0.3cm]

$\delta_{0}$ & $-0.23^{+0.69}_{-0.23}$ & $0.601^{+0.084}_{-0.077}$&$0.66^{+0.20}_{-0.14}$ & $0.223\pm 0.062$&$0.51^{+0.47}_{-0.56}$ & $0.056\pm 0.045$& $0.013^{+0.040}_{-0.055}$\\[0.3cm]
			
$\gamma$ & $7.47^{+1.2}_{-0.58}$ & $5.22^{+0.42}_{-0.47}$&$5.38^{+0.47}_{-0.72}$ & $4.58^{+0.17}_{-0.23}$ & $6.54\pm 0.59$ & $5.49\pm 0.20$& $5.97^{+0.12}_{-0.14}$\\[0.5ex]
			
\hline
\end{tabular}}
\label{tab:3}
\end{table*}

The results of statistical analysis (observational constraints) on the varying coupling model parameters for different data sets and their joint combination at $69.8\%$ CL, are listed in Table~\ref{tab:3}. Again, let us first study the coupling parameter $\delta_{0}$ (see Eq. (\ref{eq31})). In this regard, we compare results of Table~\ref{tab:3} with those of Table~\ref{tab:2}. While our analysis for constant coupling , see Table~\ref{tab:2}, indicate that except for OHD data, there is no significant evidence for DM-DE interaction, as could be seen from Table~\ref{tab:3}, for varying coupling there is considerable value for coupling $\delta_{0}$ unless we apply HST (R19) prior \cite{Riess/2019}. For OHD alone, we obtain $\delta_{0}=-0.23^{+0.69}_{-0.23}$ which means that there is an energy flow from DE to DM. However, when we combine other data to OHD, the coupling is positive, i.e., there is energy transfer from DM to DE. Our estimations show that for OHD+BAO+SNIa+CMB(+R19), the coupling is $\delta_{0}=0.056\pm 0.045 (0.013^{+0.040}_{-0.055})$, which is similar to the constant coupling model, and do not lead to any evidence for $\delta_{0}\neq 0$. In this case, we observe that except when we use the prior $H_{0}=74.02\pm1.42$ obtained from Hubble Space Telescope \cite{Riess/2019}, all data sets do not provide stringent constraints on coupling $\delta_{0}$. This result is in agreement with what is argued in Ref. \cite{Guo/2007} for BAO data. We have shown $1\sigma-3\sigma$ contour plots of ($H_{0}-\delta$), ($H_{0}-\omega^{X}$), ($\delta-\omega^{X}$) and ($\delta-\gamma$) pairs for different data sets in Figures~\ref{fig4a}, \ref{fig4b}, \ref{fig4c} \& \ref{fig4d} respectively. From Table~\ref{tab:3}, we observe that unlike constant coupling model, in the varying coupling model, not only for individual data sets but also for their joint combination, the estimated value of $H_{0}$ is not compatible with those reported by Hubble space telescope (HST) and large scale structure (LSS) experiments. This is true even when we use $H_{0}$ prior from HST \cite{Riess/2019}. However, in this case, the obtained value for Hubble constant is $H_{0}=68.9\pm 1.4$ for OHD, $H_{0}=67.43\pm 0.98$ for OHD+CMB+BAO+R19, and $H_{0}=68.69\pm 0.76(69.73\pm 0.67)$ for OHD+CMB+BAO+SNIa(+R19), which are in excellent agreement with those obtained  by Chen \& Ratra ($68 \pm 2.8$) \cite{Chen/2011}, Aubourg et al ($67.3 \pm 1.1$) \cite{Aubourg/2015}, Chen et al ($68.4^{+2.9}_{-3.3}$) \cite{Chen/2017}, Aghanim et al ($67.66 \pm 0.42$) \cite{Aghanim/2018a}, and 9-years WMAP mission ($68.92^{+0.94}_{-0.95}$) \cite{Hinshaw/2013}. Figure~\ref{fig5} depicts constraints on the Hubble constant for all combinations of data sets
considered in this work at $68\%$ CL. Our estimations show that, at $1\sigma$ error, for OHD data alone and the joint OHD+CMB+BAO+SNIa data, the dark energy EOS parameter $\omega^{X}$ varies in the phantom region, i.e., $\omega^{X}<-1$. For joint combination of all data sets plus R19, the EOS parameter is bounded in the interval $-1.030<\omega^{X}<-0.943$, which corresponds to the cosmological constant scenario. For other data sets, from Table~\ref{tab:3}, dark energy EOS parameter is varying in quintessence region, i.e., $-1<\omega^{X}<-1/3$. From Figure~\ref{fig6}, except for joint OHD+CMB+BAO+SNIa data, we observe that there is a negative correlation between Hubble constant and dark energy EOS parameter, i.e., a high value of $H_{0}$ requires a low value of $\omega^{X}$. From this figure, it could also be seen that when we constrain our model with the joint data set plus R19, there is a weak negative correlation between coupling $\delta_{0}$ and $\omega^{X}$, i.e., low values of EOS parameter require higher values of coupling constant. For these data, the DM and DE energy densities are obtained as $\Omega^{m}=0.3142^{+0.0076}_{-0.012}$ \& $\Omega^{X}=0.686^{+0.012}_{-0.0076}$, respectively. These estimations are in excellent agreement with those reported by Planck 2018 collaboration \cite{Aghanim/2018a} ($\Omega_{m}=0.3103 \pm 0.0057, \Omega_{\Lambda}=0.6897 \pm 0.0057$). From Figure~\ref{fig6}, it is obvious that there is a notable negative correlation between $H_{0}$ and $\Omega^{m}$ when we use the joint data set plus R19 to put constraints on our model. As we mentioned before in \ref{subsec:2}, in case when there is no interaction between DM and DE, $\gamma=3(1-\omega^{X})$ which in turn for cosmological constant scenario, $\omega^{X}=-1$, gives $\gamma=6$. From Table~\ref{tab:3} we can see that our estimations from the OHD+CMB+BAO+SNIa+R19 data gives $\gamma=5.97^{+0.12}_{-0.14}$, which is in the same line with $\omega^{X}=-0.990^{+0.047}_{-0.040}$ \& $\delta_{0}=0.013^{+0.040}_{-0.055}$ for the varying coupling model. Figure~\ref{fig5} depicts whisker plot with the $1\sigma$ confidence  level on the Hubble constant for varying coupling case

\section{Concluding Remarks}
\label{summary}
In this paper, we have used OHD, CMB, BAO, SNIa data and a Gaussian prior on the Hubble parameter $H_{0}$ to place observational constraints on the coupling between dark energy and dark matter in an anisotropic BI universe. We have considered two dark energy scenarios (i) constant coupling and (ii) varying coupling. The fit of both models to the joint combination of all data sets plus Gaussian prior for Hubble constant show that $\omega^{X}\approx -1$ \& $\delta(\delta_{0})\approx 0$ which indicate that the observational data favor $\Lambda$CDM model with uncoupled dark components. Also we have estimated $H_{0}=73.9^{+1.5}_{-0.95}$  and $69.73\pm 0.67$ for constant and varying coupling models, respectively. Comparing these results with the ones reported by Riess et al. (HST) \cite{Riess/2019} and the ones measured from the large scale structure (LSS) experiments \cite{Abbott/2018,Troxel/2018}, we have found that in constant coupling model there is no any significant tension for the present Hubble constant $H_{0}$. This result is important since most of the recent studies (for instance see \cite{Kumar/2019} and references therein) show that there should be a coupling (positive or negative) between the dark sector ingredients in order to alleviate the tension on $H_{0}$ . However, our statistical analyses show that in varying coupling model, the Hubble constant tension does not relax. Here, we have shown that in case of constant coupling $\delta$, a high value of $H_{0}$ (which is in high agreement with HST \& LSS) could be obtained for low value of the coupling parameter $\delta$.\\
\section*{Acknowledgment}
We are very grateful to the honorable reviewer for the illuminating suggestions that have significantly improved this paper in terms of research quality and presentation. The authors (A. K. Y. and N. A.) express their gratitude to the Deanship of Scientific Research at King Khalid University for funding this work through the Research Group Program under Grant No. RGP. 2/195/43.

\end{document}